\documentclass[preprint,showpacs,preprintnumbers,amsmath,amssymb]{revtex4}

\usepackage{graphicx}
\usepackage{epsfig}
\usepackage{amsmath}
\usepackage{graphics}

\begin{document}

\title{Tunneling properties of Bogoliubov mode and spin wave modes in supercurrent states 
of a spin-1 ferromagnetic spinor Bose--Einstein condensate} 
\author{Shohei Watabe$^{1,2}$, Yusuke Kato$^{3}$, and Yoji Ohashi$^{1,2}$} 
\affiliation{$^{1}$ 
Department of Physics, Keio University, 3-14-1 Hiyoshi, Kohoku-ku, Yokohama 223-8522, Japan}
\affiliation{$^{2}$ CREST(JST), 4-1-8 Honcho, Kawaguchi, Saitama 332-0012, Japan}
\affiliation{$^{3}$ Department of Basic Science, The University of Tokyo 153-8902, Japan}

\begin{abstract} 
We investigate tunneling properties of collective excitations in the ferromagnetic phase of a spin-1 spinor Bose--Einstein condensate (BEC). In addition to the Bogoliubov mode, this superfluid phase has two spin excitations, namely, the gapless transverse spin wave and the quadrupolar mode with a finite excitation gap. In the mean-field theory at $T=0$, we examine how these collective modes tunnel through a barrier potential that couples to the local density of particles. 
In the presence of supercurrent with a finite momentum $q$, while the Bogoliubov mode shows the so-called anomalous tunneling behavior (which is characterized by perfect transmission) in the low energy limit, 
the transverse spin-wave transmits perfectly only when the momentum $k$ of this mode coincides with $\pm q$. 
At $k=\pm q$, the wave function of this spin wave has the same form as the condensate wave function in the current carrying state, so that the mechanism of this perfect transmission is found to be the same as tunneling of supercurrent. 
Using this fact, the perfect transmission of the spin wave is proved for a generic barrier potential. 
We show that such perfect transmission does not occur in 
the quadrupolar mode. Further, we consider the effects of potentials breaking U(1) and spin rotation symmetries on  the transmission  properties of excitations. Our results would be useful for understanding excitation properties of spinor BECs,
as well as the anomalous tunneling phenomenon in Bose superfluids.
\end{abstract}

\maketitle


\section{Introduction}\label{SecI}
\par
Recently, an anomalous tunneling phenomenon has been extensively discussed in the field of Bose gas superfluids~\cite{Kovrizhin2001,Kagan2003,Danshita2006,FujitaMThesis,Kato2008,Watabe2008,Tsuchiya2008,Ohashi2008,Watabe2009RefleRefra,Takahashi2009,Tsuchiya2009,WatabeKato2010,Takahashi2010,WatabeDthesis,WatabeKatoLett,WatabeKatoFull}. 
In this phenomenon, the Bogoliubov mode tunnels through a barrier without reflection in the low energy limit. 
This tunneling property is quite different from that of an ordinary single particle in quantum mechanics, where {\it perfect reflection} occurs in the low energy limit. 
The anomalous tunneling phenomenon occurs even in the presence of a finite superflow, except in the critical supercurrent state. 
In the critical current state, tunneling of the Bogoliubov mode is accompanied by a finite reflection in the low energy limit~\cite{Danshita2006,Ohashi2008,Takahashi2009}. Since the Bogoliubov mode is a collective Nambu-Goldstone (NG) mode associated with the broken U(1) gauge symmetry in superfluid phases, the anomalous tunneling phenomenon may be considered as a fundamental property that Bose superfluids generally have.
\par
While the Bogoliubov mode dominates over low energy properties of a (spinless) scalar Bose--Einstein condensate (BEC), a spinor BEC also has other collective modes associated with spin degrees of freedom. 
It is an interesting problem whether or not such spin wave modes also exhibit anomalous tunneling behaviors as in the case of the Bogoliubov mode. In particular, in the ferromagnetic spinor BEC~\cite{Ohmi1998,Ho1998}, in addition to the ordinary Bogoliubov mode, there is a gapless transverse spin wave mode with a quadratic dispersion $E\propto p^2$ (where $p$ is the momentum of the spin wave), which is similar to the magnon in a ferromagnet~\cite{Anderson1997}. 
Thus, using the ferromagnetic spinor BEC, one can conveniently study physical properties of both the type-I Nambu-Goldstone mode (Bogoliubov mode) characterized by $E\propto p^{2n+1}$ and type-II Nambu-Goldstone mode (transverse spin wave) characterized by $E\propto p^{2n}$, where $n$ is an integer~\cite{Nielsen1976}. We also note that the ferromagnetic phase of a spin-1 BEC has the other spin mode, called the quadrupolar mode, with a finite excitation gap. Thus, this superfluid phase is a very useful system to examine how the detailed excitation spectra of collective excitations affect their tunneling properties. 
\par
In this paper, we investigate tunneling properties of low energy collective excitations in the ferromagnetic phase of a spin-1 spinor BEC at $T=0$. In a previous paper~\cite{WatabeKato2010,WatabeDthesis,WatabeKatoLett,WatabeKatoFull}, two of the authors showed the anomalous tunneling behavior of the transverse spin wave in the absence of superflow. In this paper, we extend Refs.~\cite{WatabeKato2010,WatabeDthesis,WatabeKatoLett,WatabeKatoFull} to the case with a finite superflow. Within the framework of the mean-field theory for the spin-1 BEC at $T=0$, we determine the spatial variation of condensate wave functions around the barrier by solving the Gross--Pitaevskii (GP) equation. We then clarify tunneling properties of collective excitations by solving the Bogoliubov equations. 
We show that the low-energy tunneling properties of spin wave excitations are very different from those of the Bogoliubov mode in the presence of a finite supercurrent. In particular, perfect transmission of the transverse spin wave is shown to occur through a generic potential barrier, not in the low momentum limit, but in the case 
when the magnitude of spin wave momentum $|k|$ and that of the momentum of the supercurrent $|q|$ are equal. 
(Note that perfect transmission of the Bogoliubov mode always occurs in the zero momentum limit ($k\to 0$), irrespective of the value of $q$.) 
\par
This paper is organized as follows. In Sec.~\ref{SecII}, we explain our formulation for tunneling of collective excitations through a barrier 
in current carrying states of a spin-1 ferromagnetic spinor BEC. 
We discuss tunneling properties of the transverse spin wave and the quadrupolar mode in Secs.~\ref{SecIII} and \ref{SecIV}, respectively. 
In Sec.~\ref{SecV}, we examine how the symmetry of a barrier potential affects the anomalous tunneling properties of the Bogoliubov mode and the transverse spin wave. 

\begin{figure}[tbp]
\begin{center}
\includegraphics[width=10cm]{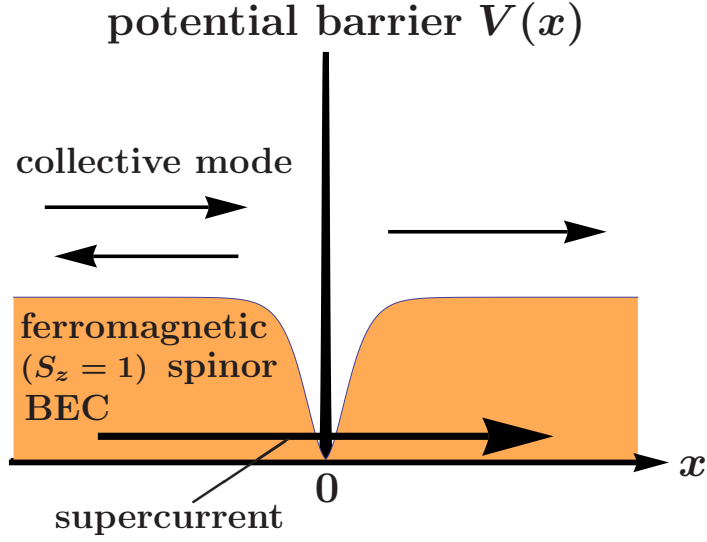}
\end{center}
\caption{(color online) Model of a spin-1 ferromagnetic spinor BEC with a potential barrier $V(x)$ put around $x=0$. The ferromagnetic condensate wave function is described by $S_z=1$ spin component, and a finite supercurrent flows in the $x$-direction. In this paper, a collective mode is injected from $x=-\infty$. For simplicity, we assume a one-dimensional system and ignore the unimportant $y$- and $z$-direction.}
\label{fig1}
\end{figure} 


\section{Mean-field theory of spin-1 BEC}\label{SecII} 
\par
We consider the supercurrent state of a ferromagnetic spin-1 spinor BEC with a barrier potential put around $x=0$, as schematically shown in Fig.~\ref{fig1}. Since the $y$- and $z$-direction are not important in this paper, we simply assume a one-dimensional system. The spin quantization axis is taken parallel to the $z$-direction, and we introduce a three-component condensate wave function
\begin{equation}
{\hat \Phi}(x,t)=
\left(
\begin{array}{c}
\Phi_{+1}(x,t)\\
\Phi_0(x,t)\\
\Phi_{-1}(x,t)\\
\end{array}
\right),
\label{eq.1}
\end{equation}
where $\Phi_j(x,t)$ is the component of magnetic sublevel $j=S_z$ ($\pm 1$ and $0$). As usual, the condensate wave function ${\hat \Phi}(x,t)$ obeys the GP equation obtained from the variational principle of the action
\begin{equation}
I=\int dt \int dx {\mathcal L}.
\label{eq.2}
\end{equation}
Here, ${\mathcal L}=i\hbar {\hat \Phi}^\dagger (x,t)\partial_t{\hat \Phi}(x,t)-{\mathcal H}(x,t)$ is the Lagrangian density, where the Hamiltonian density ${\mathcal H}$ is given by~\cite{Ohmi1998,Ho1998} 
\begin{equation}
{\mathcal H}
= 
-
{\hbar^2 \over 2m} 
{\hat \Phi}^\dagger(x,t)
\partial_x^{2}
{\hat \Phi}(x,t)
+V(x) \rho (x,t) 
+
{c_{0} \over 2} \rho^{2} (x,t) 
+ 
{c_{1} \over 2} {\bf F}^{2} (x,t)
-
g\mu_{\rm B}BF_z(x,t). 
\label{eq.2b}
\end{equation} 
Here, $V(x)$ is a barrier potential put around $x=0$ in Fig.~\ref{fig1}. $c_0=4\pi\hbar^{2}(a_{0}+2a_{2})/(3m)$ and $c_1=4\pi\hbar^{2}(a_{2}-a_{0})/(3m)$ describe, respectively, 
coupling constants of 
a spin-independent and spin-dependent interactions~\cite{Ho1998}, 
where $a_0$ and $a_2$ are the $s$-wave scattering lengths between two atoms in the total spin $S=0$ state and $S=2$ state, respectively. 
$\rho(x,t)={\hat \Phi}^\dagger(x,t){\hat \Phi}(x,t)$ is the particle density, 
and ${\bf F}= {}^{\rm t}(F_x,F_y,F_z)$ is the spin density, where $F_j(x,t)={\hat \Phi}^\dagger(x,t){\hat S}_j{\hat \Phi}(x,t)$. 
Since we are taking the spin quantization axis parallel to the $z$-direction, the $S=1$ spin matrices ${\hat S}_j (j=x,y,z)$ are given by
\begin{align}
{\hat S}_x = 
\frac{1}{\sqrt{2}}
\begin{pmatrix}
0&1&0 \\
1&0&1 \\
0&1&0
\end{pmatrix},~~ 
{\hat S}_y= 
\frac{i}{\sqrt{2}}
\begin{pmatrix}
0&-1&0 \\
1&0&-1 \\
0&1&0
\end{pmatrix}, ~~
{\hat S}_z= 
\begin{pmatrix}
1&0&0 \\
0&0&0 \\
0&0&-1
\end{pmatrix}.  
\label{eq.3}
\end{align} 
The last term in Eq. (\ref{eq.2b}) describes effects of a magnetic field $B$ applied parallel to the $z$-axis, where $g$ and $\mu_{\rm B}$ are the Land\'e's $g$ factor and Bohr magneton, respectively. We take $B$ to be non-negative without loss of generality. 
\par
Taking the variation of the action $I$ in Eq.~(\ref{eq.2}) 
with respect to ${\hat \Phi}^\dagger(x,t)$, we obtain the time-dependent GP equation, 
\begin{eqnarray}
i\hbar
{\partial {\hat \Phi}(x,t) \over \partial t}
=\left(
\begin{array}{ccc}
h(x,t)+c_1F_z-g\mu_{\rm B}B& \displaystyle{c_1 \over \sqrt{2}} F_-& 0\\
\displaystyle{c_1 \over \sqrt{2}}F_+ & h(x,t) & \displaystyle{c_1 \over \sqrt{2}}F_-\\
0 & \displaystyle{c_1 \over \sqrt{2}}F_+ & h(x,t)-c_1F_z+g\mu_{\rm B}B \\
\end{array}
\right)
{\hat \Phi}(x,t).
\label{eq.4}
\end{eqnarray}
Here, $h(x,t)=- \hbar^{2}\partial_x^2/(2m) + V(x) + c_{0} \rho(x,t)$, and $F_{\pm}=F_{x} \pm i F_{y}$. The time-independent GP equation for 
stationary states is then obtained by setting $\Phi_j(x,t)=e^{-i\mu_jt/\hbar}\Phi_j(x)$ under the condition $2\mu_0=\mu_1+\mu_{-1}$~\cite{Nistazakis2007}. In this paper, taking $\mu_{\pm1}=\mu_0=\mu$~\cite{Nistazakis2007,NoteChemical}, we obtain 
\begin{eqnarray}
\left(
\begin{array}{ccc}
h(x)+c_1F_z-g\mu_{\rm B}B& \displaystyle{c_1 \over \sqrt{2}} F_-& 0\\
\displaystyle{c_1 \over \sqrt{2}}F_+ & h(x) & \displaystyle{c_1 \over \sqrt{2}}F_{-}\\
0 & \displaystyle{c_1 \over \sqrt{2}}F_+ & h(x)-c_1F_z+g\mu_{\rm B}B \\
\end{array}
\right)
{\hat \Phi}(x)=0,
\label{eq.5}
\end{eqnarray}
where $h(x)=- \hbar^{2}\partial_x^2/(2m) -\mu + V(x) + c_{0} \rho(x)$.
\par
The ferromagnetic state is always realized as the ground state when the spin-spin interaction 
is ferromagnetic ($c_1<0$) and $B=0$. This region in $c_1$-$B$ plane becomes wider as $c_{1} < g\mu_{\rm B}B / \rho_{0}$. In the ferromagnetic state, all the Bose atoms occupy the $S_z=1$ state as \begin{equation}
{\hat \Phi}(x)=
\left(
\begin{array}{c}
\Phi_{+1}(x)\\
0\\
0\\
\end{array}
\right).
\label{eq.6}
\end{equation}
In the uniform system ($V(x)=0$), the $S_z=1$ component in the supercurrent state has the form $\Phi_{+1}(x)=\sqrt{\rho_0}e^{iqx}$, and the chemical potential $\mu$ is given by $\mu=c_+\rho_0+\hbar^2q^2/(2m)-g\mu_{\rm B}B$. Here, $\rho_0$ is the particle density of the uniform Bose gas, and $c_+=c_0+c_1$ is 
the coupling constant of the interaction between the atoms in the $S_z=1$ state. 
In addition, as will be shown soon later, 
the other spin components, $\Phi_0(x)$ and $\Phi_{-1}(x)$, are not induced by the barrier $V(x)$. Using these results, one finds that the GP equation (\ref{eq.5}) in the ferromagnetic state reduces to
\begin{equation}
\left [
-\frac{\hbar^{2}}{2m} \frac{d^{2}}{dx^{2}}
+ V(x) 
- c_{+} \rho_0 
- \frac{\hbar^{2}q^{2}}{2m} 
+ 
c_{+}  \rho(x)
\right ] \Phi_{+1} (x) = 0. 
\label{eq.7}
\end{equation} 
We note that Eq.~(\ref{eq.7}) has the same form as the GP equation for the scalar BEC, when we regard $c_+$ as 
the coupling constant of an interaction between spinless Bose atoms.
\par

\begin{figure}[tbp]
\begin{center}
\includegraphics[width=6cm]{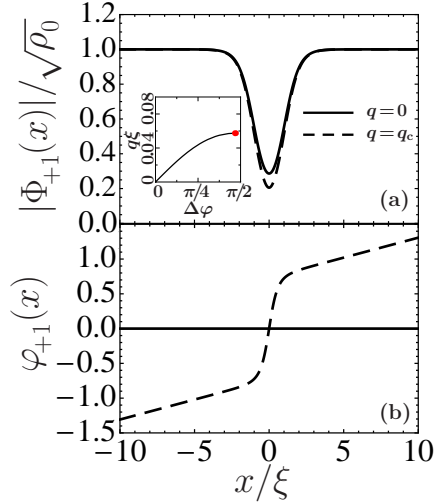}
\end{center}
\caption{(color online) Calculated condensate wave function $\Phi_{+1}(x)=|\Phi_{+1}(x)|e^{i\varphi_{+1}(x)}$ in the supercurrent state of a ferromagnetic spin-1 spinor BEC at $T=0$ 
for the barrier potential in Eq.~(\ref{eq.8}) with $V_{0} = 2 c_{+}\rho_0$.
Since $\Phi_0(x)$ and $\Phi_{-1}(x)$ are absent everywhere, we do not show them here. (a) amplitude $|\Phi_{+1}(x)|$. (b) phase $\varphi_{+1}(x)$. In panel (b), we set $\varphi_{+1}(0)=0$. 
The inset shows the momentum $q$ of the supercurrent as a function of the phase difference $\Delta \varphi$. The red point ($q=0.0574\xi^{-1}\equiv q_{\rm c}$) in the inset shows the critical current state. 
}
\label{fig2}
\end{figure} 

\par
Figure \ref{fig2} shows the spatial variation of the condensate wave function $\Phi_{+1} (x)=|\Phi_{+1} (x)|e^{i\varphi_{+1}(x)}$. In this figure, we consider the potential barrier
\begin{equation}
V(x)=V_0 e^{-x^2/\xi^2},
\label{eq.8}
\end{equation}
where $\xi=\hbar/\sqrt{mc_+\rho_0}$ is the healing length. Although we have numerically solved the GP equation (\ref{eq.5}) to include the possibility of finite $\Phi_0(x)$ and $\Phi_{-1}(x)$ components around $x=0$, these components have turned out to be absent. Thus, one may safely use the simpler GP equation (\ref{eq.7}). As expected, Fig.~\ref{fig2} (a) shows that the condensate wave function is suppressed around the potential barrier. In addition, panel (b) shows that the phase $\varphi_{+1}(x)$ of the condensate wave function spatially varies. This behavior reflects the presence of a finite superflow. 
Far away from the barrier, the phase factor reduces to the form 
$\varphi_{+1}(x\to\pm\infty)=qx+C_\pm$~\cite{Danshita2006}. 
The phase difference $\Delta\varphi\equiv C_+-C_-$ can be conveniently calculated from~\cite{Baratoff1970, Danshita2006} 
\begin{align}
\Delta \varphi \equiv q \int_{-\infty}^{\infty} 
dx 
\left ( \frac{\rho_0}{|\Phi_{+1}(x)|^{2}} - 1 \right ). 
\label{eq.9}
\end{align} 
The inset of 
Fig.~\ref{fig2} (a) 
shows the momentum $q$ of the supercurrent as a function of $\Delta\varphi$. 
As in the case of the scalar BEC, the upper critical momentum $q_{\rm c}$ exists at $\Delta\varphi\simeq\pi/2$ 
(the red point in the inset).
\par
Once the condensate wave function ${\hat \Phi}(x)$ is determined, collective excitations are conveniently obtained by considering small fluctuations around ${\hat \Phi}(x)$. Setting ${\hat \Phi}(x,t)={\hat \Phi(x)}+{\hat \phi}(x,t)$ (${\hat \phi}= {}^{\rm t}(\phi_{+1},\phi_0,\phi_{-1})$), and substituting it into Eq. (\ref{eq.4}), one obtains the Bogoliubov equations~\cite{note},
\begin{align}
i\hbar \frac{\partial \phi_{\pm 1}}{\partial t} 
= &  
\left [ 
h(x) \mp g\mu_{\rm B}B + R_{\pm 1,\pm 1}'^{(+)} + c_{1}( \pm  F_{z} + |\Phi_{0}|^{2}) \right ]\phi_{\pm 1} 
+ 
R_{\pm1,\pm1}^{(+)} \phi_{\pm 1}^* 
\nonumber 
\\
&
+ 
P_{\pm 1}\phi_{0}+R_{0,\pm1}^{(+)}  \phi_{0}^* 
+ 
R_{\mp 1,\pm 1}'^{(-)} \phi_{\mp 1} 
+ 
( R_{+1,-1}^{(-)}  + c_{1} \Phi_{0}^{2} ) \phi_{\mp 1}^*, 
\label{eq.10}
\\
i\hbar \frac{\partial \phi_{0}}{\partial t} 
= & 
\left ( h(x) 
+ c_{1} \rho (x) + R_{0,0}'^{(-)}\right ) \phi_{0} 
+ 
( c_{0} \Phi_{0}^{2} + 2c_{1} \Phi_{+1} \Phi_{-1} ) \phi_{0}^* 
\nonumber 
\\
&
+ \sum\limits_{j=\pm 1}(P_{j}^{*} \phi_{j} + R_{0,j}^{(+)} \phi_{j}^* ), 
\label{eq.11}
\end{align}
where $P_{\pm 1} \equiv  ( c_{0} + c_{1} ) \Phi_{0}^{*}\Phi_{\pm 1} + 2 c_{1}  \Phi_{\mp 1}^{*}\Phi_{0}$, $R_{i,j}^{(\pm)} \equiv (c_{0}\pm c_{1})\Phi_{i} \Phi_{j}$, and $R_{i,j}'^{(\pm)} \equiv (c_{0}\pm c_{1})\Phi_{i}^{*} \Phi_{j}$. 
In obtaining Eqs. (\ref{eq.10}) and (\ref{eq.11}), we have retained terms 
up to $O({\hat \phi}(x))$. 
In the ferromagnetic case, Eqs. (\ref{eq.10}) and (\ref{eq.11}) reduce to
\begin{align}
i\hbar \frac{\partial \phi_{+1}}{\partial t} 
& = 
\left [
-\frac{\hbar^{2}}{2m} \frac{d^{2}}{dx^{2}}
+ V (x)- c_{+} \rho_0 - \frac{\hbar^{2}q^{2}}{2m} 
+ 2 c_{+} \rho(x) 
\right ] \phi_{+1}
+ 
c_{+} \Phi_{+1}^{2}(x) \phi_{+1}^{*},  
\label{eq.12} 
\\
i\hbar \frac{\partial \phi_{0}}{\partial t} 
& = 
\left [
-\frac{\hbar^{2}}{2m} \frac{d^{2}}{dx^{2}}+ V (x)
- c_{+} \rho_0- \frac{\hbar^{2}q^{2}}{2m} + c_{+} \rho(x)+g\mu_{\rm B}B 
\right ] \phi_{0}, 
\label{eq.13}
\\
i\hbar \frac{\partial \phi_{-1}}{\partial t} 
& = 
\left [
-\frac{\hbar^{2}}{2m} \frac{d^{2}}{dx^{2}}
+ V (x)-c_{+} \rho_0-\frac{\hbar^{2}q^{2}}{2m}
+(c_{+}-2 c_{1})\rho(x)+2g\mu_{\rm B}B
\right ] \phi_{-1}. 
\label{eq.14}
\end{align}
Since Eqs. (\ref{eq.12})-(\ref{eq.14}) are independent, 
we can treat each component $\phi_j(x)$ separately. 
Equation~(\ref{eq.12}) has the same form as the ordinary 
(time-dependent) Bogoliubov equation 
in the scalar BEC~\cite{note}. 
We thus see that the mode described by $\phi_{+1}(x)$ is the Bogoliubov mode in the present ferromagnetic spinor BEC and exhibits the anomalous tunneling phenomenon as in the case of the scalar BEC~\cite{Kovrizhin2001, Kagan2003, Danshita2006}.  
Namely, the Bogoliubov mode in the low energy limit always tunnels through the barrier without reflection in the supercurrent state~\cite{Danshita2006,Ohashi2008}, except in the critical current state. 
In the critical supercurrent state, the Bogoliubov mode tunnels through the barrier with a finite reflection~\cite{Danshita2006,Ohashi2008,Takahashi2009}.
\par
To identify roles of $\phi_0$ and $\phi_{-1}$, it is convenient to consider the uniform case with no external magnetic field ($V(x)=B=0$) and no supercurrent ($q=0$). In this simple case, because of $\rho(x)=\rho_0$, 
Eq.~(\ref{eq.13}) gives the gapless spectrum with quadratic dispersion $E_0=p^2/(2m)$. This means that $\phi_0$ describes the ferromagnetic spin wave mode, which we call the transverse spin wave~\cite{Ho1998}. In addition, 
Eq.~(\ref{eq.14}) gives $E_{-1}=p^2/(2m)+2|c_1|\rho_0$ with a finite excitation gap $2|c_1|\rho_0$. This mode is called the quadrupolar mode~\cite{Ho1998}. The transverse spin wave described by $\phi_0$ changes the total magnetization $M$ by $g\mu_{\rm B}$. On the other hand, the quadrupolar mode described by $\phi_{-1}$ changes $M$ by $2g\mu_{\rm B}$. 
\par

The current-flowing ferromagnetic state in the uniform case is stable only when the excitation energies of both the transverse spin wave 
\begin{equation}
E_0 =\frac{p^2}{2m}-\frac{\hbar^2 q^2}{2m}+g\mu_{\rm B}B
\end{equation}
and the quadrupolar spin wave 
\begin{equation}
E_{-1} =\frac{p^2}{2m}-\frac{\hbar^2 q^2}{2m}+2|c_1|\rho_0+2g\mu_{\rm B}B
\end{equation}
are non-negative for arbitrary $p$. Since $E_0<E_{-1}$, the stability condition of the current-flowing ferromagnetic state is given by
\begin{equation}
\frac{\hbar^2 q^2}{2m}\le g\mu_{\rm B}B.
\label{eq: stability}
\end{equation}
In the following, we consider the cases where Eq.~(\ref{eq: stability}) holds.
\par
\section{Tunneling properties of transverse spin wave}
\label{SecIII}
\par
To examine the stationary solution of the transverse spin wave described by $\phi_0(x,t)$, we set $\phi_0(x,t)=e^{-iEt/\hbar}\phi_0(x)$ in Eq. (\ref{eq.13}). 
We then have
\begin{align}
E {\phi}_{0} (x)
=\left [
-\frac{1}{2} \frac{d^{2}}{dx^{2}}+ V (x) - 1 - \frac{q^{2}}{2} 
+ |\Phi_{+1} (x)|^{2}
+ \Omega
\right ] {\phi}_{0} (x). 
\label{eq.15}
\end{align} 
In obtaining Eq.~(\ref{eq.15}), we have introduced the dimensionless variables, $\bar{E} \equiv E/(c_{+}\rho_{0})$, $\bar{x} \equiv x/\xi$, $\bar{q} = q\xi$, $\bar{c}_{i} \equiv c_{i}/c_{+}$ ($i=1,2$), $\bar{V} \equiv V / (c_{+}\rho_0)$, $\bar{\Phi}_{+1} \equiv \Phi_{+1}/\sqrt{\rho_0}$, $\bar{\phi}_{+1} \equiv \phi_{+1}/\sqrt{\rho_0}$, $\Omega \equiv g \mu_{\rm B} B/(c_{+}\rho_0)$, and $\bar{t} \equiv t c_{+} \rho_0 / \hbar$, and have omitted the bars from these rescaled variables, for simplicity. We always use these dimensionless units in the following discussions. 
\par
In the uniform system, Eq. (\ref{eq.15}) has the plain wave solution $\phi_{0}(x)=e^{\pm i k x}$ with the mode energy
\begin{equation}
E = \frac{1}{2} ( k^{2} - q^{2} ) + \Omega.
\label{eq.16}
\end{equation} 
Note that the stability condition Eq.~(\ref{eq: stability}) of the ground state is expressed as $2\Omega-q^2\ge 0$ in the dimensionless units.
For a given excitation energy $E$, the spin wave momentum far away from the barrier is given by $k = \sqrt{ 2  (E - \Omega) + q^{2} }$. 

We prove that perfect transmission of the transverse spin wave occurs at $|k|=q$  for a generic barrier potential $V(x)$. 
With use of (\ref{eq.16}), we first note that Eq. (\ref{eq.15}) reduces to 
\begin{align}
0 
& = 
\left [
-\frac{1}{2} \frac{d^{2}}{dx^{2}}
+ V (x)
- 1 - \frac{q^{2}}{2}
+ |\Phi_{+1} (x)|^{2} 
\right ] 
 {\phi}_{0} (x)
\label{eq.27}
\end{align} 
when $k=q$. When ${\phi}_{0} (x)$ is replaced by the ferromagnetic condensate wave function $\Phi_{+1}(x)=\Phi_{+1}(x,q)$ with the momentum $q$, Eq.~(\ref{eq.27}) coincides with the GP equation,
\begin{align}
0 
& = 
\left [
-\frac{1}{2} \frac{d^{2}}{dx^{2}}
+ V (x)
- 1 - \frac{q^{2}}{2}
+ |\Phi_{+1} (x,q)|^{2} 
\right ]
 {\Phi}_{+1} (x,q), 
\label{eq: eq.7-dimensionless}
\end{align} 
which is nothing but Eq.~(\ref{eq.7}) in terms of the dimensionless units introduced previously. We also note that when ${\phi}_{0} (x)$ is replaced by $\Phi_{+1}^{*} (x,q)$, Eq.~(\ref{eq.27}) coincides with the complex conjugation of Eq.~(\ref{eq: eq.7-dimensionless}).  It immediately follows that $\Phi_{+1}(x,q)$ and $\Phi_{+1}^{*}(x,q)$ are linearly independent solutions to Eq. (\ref{eq.27}). 
The general solution to Eq. (\ref{eq.27}) is then expressed as
\begin{align}
{\phi}_{0}(x) = 
& \alpha \Phi_{+1} (x,q) + \beta \Phi_{+1}^{*} (x,q), 
\label{eq.28}
\end{align} 
where the coefficients $\alpha$ and $\beta$ are determined so that (\ref{eq.28}) satisfies the boundary conditions of the tunneling problem
\begin{eqnarray}
\begin{array}{ll}
{\phi}_{0} = e^{ikx} + R_0 e^{-ikx}, &  (x \ll -1),\\ 
{\phi}_{0} = T_0 e^{ikx},  & (x \gg 1), 
\end{array}
\label{eq.32}
\end{eqnarray}
for $k=q$.
When we set the phase of $\Phi_{+1}(x,q)$ so as to satisfy $\varphi_{+1}(x=0)=0$, the asymptotic form of the condensate wave function may be written as $\Phi_{+1}(x,q) \to e^{i [qx + {\rm sgn} (x) \Delta \varphi / 2 ]}$ ($|x| \gg 1$)~\cite{NoteFunc}, where $\Delta\varphi$ is the phase difference introduced in Eq. (\ref{eq.9}). From this asymptotic behavior of $\Phi_{+1}(x,q)$ and the compatibility of Eqs.~(\ref{eq.28}) and (\ref{eq.32}), one obtains $(\alpha,\beta)=(e^{i \Delta \varphi /2},0)$, i.e., 
\begin{align}
{\phi}_{0} (x) = e^{i\Delta\varphi/2} \Phi_{+1} (x,q), 
\label{eq.26}
\end{align}
and $(T_0,R_0)=(e^{i \Delta \varphi},0)$ at $k=q$. 
Perfect transmission $|T_0|^2=1$ at $k = q$ for an arbitrary barrier potential $V(x)$ thus follows. 

The same discussion is applicable to the case when $k=-q$. 
In this case, note that the condensate wave function in the supercurrent with the momentum $-q$ ($\equiv \Phi_{+1}(x,-q)$) is related to the condensate wave function with the supercurrent momentum $+q$ ($\equiv \Phi_{+1}(x,+q)$) as $\Phi_{+1}(x,-q)=\Phi_{+1}^*(x,+q)$. 
By repeating the above discussion by simply replacing $\Phi_{+1}$ by $\Phi_{+1}^{*}$, one may derive perfect transmission at $k=-q$.

The above argument applies when $q=q_{\rm c}$ and hence the transverse spin wave transmits perfectly when $q=q_{\rm c}$ and $k=q_{\rm c}$. This is contrasted with the transmission property of the Bogoliubov mode, which does not exhibit the perfect transmission when $q=q_{\rm c}$~\cite{Danshita2006,Ohashi2008,Takahashi2009}. 
\par
In the absence of supercurrent $(q=0)$, 
Eq.~(\ref{eq.26}) does not necessarily imply 
that the wave function of the transverse spin wave mode ($k=0$) has the the supercurrent behavior (i.e., the same form as the supercurrent wave function). The above derivation of perfect transmission does not apply because the incident and reflection waves are indistinguishable at $k=0$. When we discuss the tunneling problem in the limit $k\rightarrow 0$, we have to consider the problem with $k$ being small but finite, and then take the limit $k\to 0$. Following this procedure, we can show that this spin wave (with the momentum $k\ll 1$) has the same form as the supercurrent wave function with the same momentum $k$ with the accuracy of ${\cal O}(k)$ and from this fact, the perfect transmission follows in the limit $k\rightarrow 0$ when $q=0$. The derivation is given in Appendix A. To conclude, the transverse spin wave, 
as well as the Bogoliubov mode, always has the supercurrent behavior, when the perfect transmission occurs.

\par
\begin{figure}[tbp]
\begin{center}
\includegraphics[width=12cm]{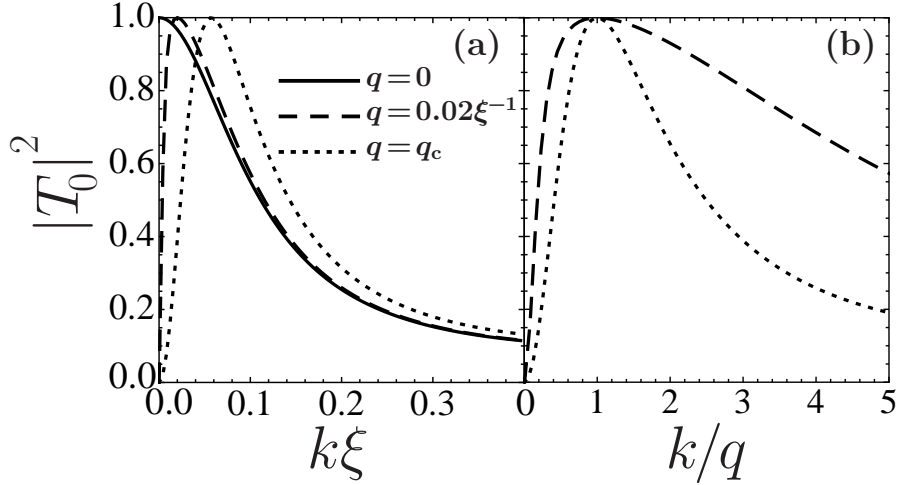}
\end{center}
\caption{
Transmission probability $|T_0|^2$ of the transverse spin wave, when the barrier is given by 
Eq.~(\ref{eq.8}) with $V_{0}=2c_{+}\rho_0$. (a) momentum dependence. (b) $|T_0|^2$ as a function of $k/q$. 
In obtaining this result, we have used the finite element method, imposing the boundary condition in Eq. (\ref{eq.32}). This method is also used to obtain Fig.~\ref{fig5}.
}
\label{fig3}
\end{figure} 
%
Let us see examples. Figure~\ref{fig3} shows transmission probability $|T_0|^2$ of the transverse spin wave in the presence of the Gaussian-type barrier Eq.~(\ref{eq.8}) with $V_{0}=2c_{+}\rho_0$. From the scaled plot (b), we see that the perfect transmission occurs at $k=q$. 

Another example is provided by the exact solution in the presence of a $\delta$-functional barrier $V(x)=V_0\delta(x)$. 
In the case, the GP equation (\ref{eq.7}) has the following solution~\cite{Hakim1997,Danshita2006}:
\begin{align}
\Phi_{+1} (x) = e^{i [ q x - {\rm sgn}(x) \theta_{q} ] } [\gamma (x) - {\rm sgn} (x) i q ], 
\label{eq.17}
\end{align}
where $\gamma(x)=\sqrt{1-q^{2}}\tanh[\sqrt{1-q^{2}}(|x|+x_{0})]$. The phase factor $\theta_q$ is given by
\begin{equation}
e^{i\theta_{q}}=
{\gamma_{q} - i q \over \sqrt{\gamma_{q}^{2}  + q^{2}}}, 
\label{eq.18}
\end{equation} 
where $\gamma_{q}$ denotes $\sqrt{1-q^{2}} \tanh(\sqrt{1-q^{2}}x_{0})$ and $x_{0}$ is determined from the boundary condition on the derivative of $\Phi_{+1}(x)$ at $x=0$, which gives
\begin{align}
V_{0} ={(1-q^{2}) \gamma_{q} - \gamma_{q}^{3} \over \gamma_{q}^{2} + q^{2}}. 
\label{eq.19}
\end{align} 
Substituting Eq.~(\ref{eq.17}) into Eq. (\ref{eq.15}), one obtains
\begin{equation} 
{\phi}_{0} (x) = [-{\rm sgn}(x)ik + \gamma(x) ] e^{ikx}. 
\label{eq.20}
\end{equation}
To solve the tunneling problem, we set 
\begin{eqnarray}
\begin{array}{ll}
{\phi}_{0} (x) = 
[ik + \gamma (x) ]e^{ikx}  + R_0[-ik + \gamma (x)] e^{- ikx}, & (x \leq 0),\\ 
{\phi}_{0} (x) = 
T_0[- ik + \gamma (x)] e^{ + ikx}, & (x \geq 0). 
\label{eq.21}
\end{array}
\end{eqnarray}
As usual, the coefficients $R_0$ and $T_0$ are determined so as to satisfy the boundary conditions, 
\begin{eqnarray}
\begin{array}{l}
{\phi}_{0} (+0)={\phi}_{0} (-0), 
\\
\displaystyle
\left. \frac{d {\phi}_{0}(x)}{dx} \right  |_{x=+0} 
=
\left. \frac{d {\phi}_{0}(x)}{dx} \right  |_{x=-0}  +  2 V_{0} {\phi}_{0}(0).
\label{eq.22}
\end{array}
\end{eqnarray} 
We then have
\begin{align}
T_0 = & 
ik \left  [ 
\frac{1}{-ik + \gamma_{q}} + \frac{ \gamma_{q} }
{-q^{4} + k \gamma_{q} (i+k \gamma_{q} ) + q^{2} (1 + k^{2} - \gamma_{q}^{2})}
\right ], 
\label{eq.23}
\\
R_0 = &
- 1 + ik 
\left [ 
- \frac{1}{-ik+\gamma_{q} } 
+ 
\frac{\gamma_{q} }
{ - q^{4} + k \gamma_{q} (i + k \gamma_{q} ) + q^{2} (1+k^{2} - \gamma_{q}^{2})}
\right ]. 
\label{eq.24}
\end{align}
\par
In the absence of a superflow ($q=0$), 
Eq.~(\ref{eq.23}) reduces to
\begin{align}
T_0 = \frac{i {\gamma}_{0} (1 + k^{2})}{(i + k {\gamma}_{0}) (-ik + {\gamma}_{0})}. 
\label{eq.25}
\end{align}
Thus, perfect transmission ($|T_0|^2=1$) is obtained in the long wavelength limit ($k\to0$)~\cite{WatabeKato2010, WatabeDthesis, WatabeKatoLett, WatabeKatoFull}. 
In contrast, Eq. (\ref{eq.23}) vanishes in the limit $k\to0$ when $q$ is finite. 
This shows clearly that, in contrast to the Bogoliubov mode, anomalous tunneling of the transverse spin is not realized in the long wavelength limit in the presence of a finite superflow. 

\begin{figure}[tbp]
\begin{center}
\includegraphics[width=12cm]{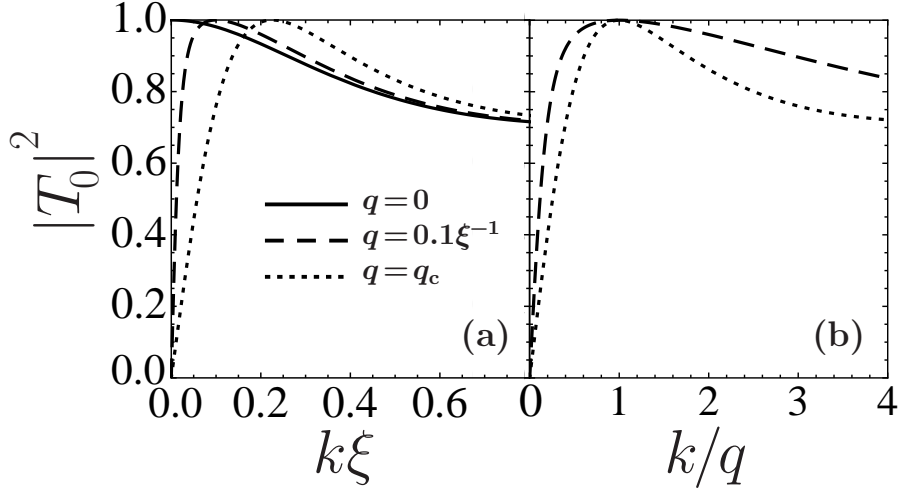}
\end{center}
\caption{(a) Transmission probability $|T_0|^2$ of the transverse spin wave in the case of a $\delta$-functional barrier. 
Panel (b) also shows $|T_0|^2$ as a function of $k/q$ to clearly show that the perfect transmission always occurs at $k=q$. We take $V_{0}=2(c_{+}n)/\xi$.}
\label{fig4}
\end{figure} 

\par
When $k=q>0$, Eq. (\ref{eq.23}) becomes
\begin{align} 
T_0 = \frac{\gamma_{q} + i q  }{\gamma_{q} - i q } \equiv e^{- 2 i \theta_{q}}, 
\label{eq.25b}
\end{align}
which gives perfect transmission $|T_0|^2=1$. In the scalar BEC, 
it has been pointed out that the wave function of the Bogoliubov mode has 
the same form as the condensate wave function 
when the anomalous tunneling occurs in the absence of supercurrent~\cite{Kato2008} and in the presence of supercurrent~\cite{Ohashi2008}. In the present case, from the comparison of Eq. (\ref{eq.17}) with Eq. (\ref{eq.21}) at $k=q$, one finds 
\begin{align}
{\phi}_{0} (x) = e^{-i\theta_{q}} \Phi_{+1} (x,q), 
\label{eq.26-2}
\end{align} 
as in the case of generic potential barrier (\ref{eq.26}). 
The fact that perfect transmission of the transverse spin wave occurs at the finite momentum $k=q$ is different from the condition of anomalous tunneling in the case of the Bogoliubov mode; however, 
the supercurrent behavior can be also seen in the transverse spin wave when perfect transmission occurs ($k=q$).
In the absence of supercurrent $(q=0)$, 
we show in Appendix B that this spin wave (with the momentum $k\ll 1$) has the same form as the supercurrent wave function with the same momentum $k$ with the accuracy of ${\cal O}(k)$. This fact illustrates the supercurrent behavior of the transverse spin wave in the limit $k\rightarrow 0$ when $q=0$.
\par
We briefly note that the amplitude transmission coefficient in Eq. (\ref{eq.23}) satisfies $T_0(-k)=T_0(k)^*$. Because of this symmetry, we find that perfect transmission also occurs at $k=-q$. 
\par
Figure~\ref{fig4} shows the transmission probability $|T_0|^2$ of the transverse spin wave through the $\delta$-functional barrier. 
When the supercurrent momentum $q$ becomes finite, panel (a) shows that the value of $k$ at which perfect transmission ($|T_0|^2=1$) is realized becomes large. At perfect transmission, panel (b) confirms that $|k|/q=1$ is satisfied, as expected. Figure \ref{fig4} also shows that the perfect transmission occurs only when $|k| = q$. 
\par 

\begin{figure}[tbp]
\begin{center}
\includegraphics[width=6cm]{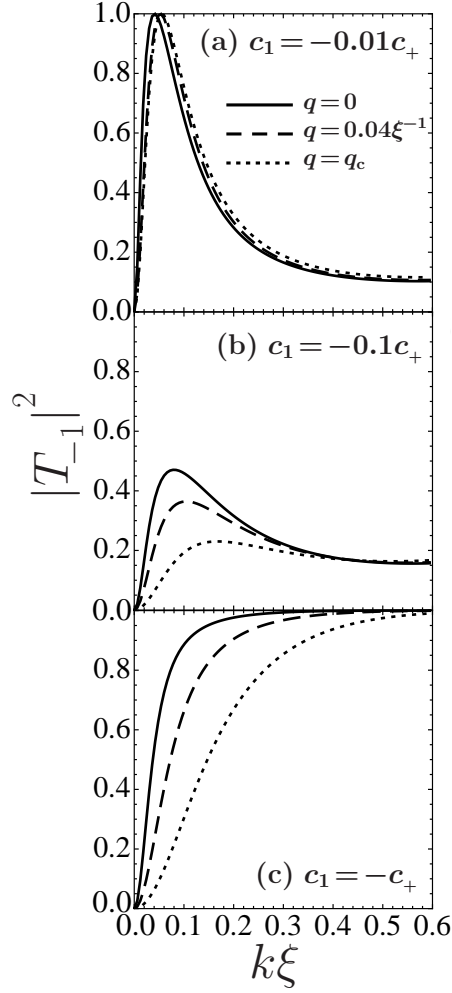}
\end{center}
\caption{Transmission probability $|T_{-1}|^2$ of the quadrupolar mode. 
The barrier potential is given by Eq.~(\ref{eq.8}) with $V_{0}=2c_{+}\rho_0$.
The ratios $c_1/c_+$ of the coupling constants are (a) -0.01, (b) -0.1, and (c) -1. 
}
\label{fig5}
\end{figure} 
\par

\section{Transmission properties of quadrupolar mode}\label{SecIV}

The quadrupolar spin wave is described by the fluctuation component $\phi_{-1}$, which obeys Eq.~(\ref{eq.14}). In the stationary state, $\phi_{-1} (x, t) = {\phi}_{-1} (x) e^{-iEt}$, this equation reduces to, in the dimensionless form,
\begin{align}
E {\phi}_{-1}
& = 
\left [
-\frac{1}{2} \frac{d^{2}}{dx^{2}}
+ V(x) 
-1 - \frac{q^{2}}{2} + (1 - 2c_{1}) |\Phi_{+1}|^{2}+ 2\Omega
\right ] {\phi}_{-1}.  
\label{eq.36}
\end{align} 
The mode energy in the uniform system is obtained as $E = (k^{2}-q^{2})/2 + 2 |c_{1}| + 2 \Omega$, which is positive for arbitrary $k$ 
when the stability condition (\ref{eq: stability}) ($2\Omega >q^2$ in dimensionless units) of the current-flowing ground state is satisfied. 
Under this condition, the momentum $k$ has the form 
$k = \sqrt{ {2} \left [ E + 2 (c_{1} - \Omega)\right ] + q^{2} }$. 
In considering the tunneling problem, the boundary conditions far away from the barrier are given by
\begin{eqnarray}
\begin{array}{ll}
{\phi}_{-1} (x) = \exp{(i k x)} + R_{-1} \exp{(- i k x)}, &  ~~~(x \ll -1),
\\
{\phi}_{-1} (x) = T_{-1} \exp{( i k x)}, & ~~~(x \gg + 1). \\
\label{eq.37}
\end{array}
\end{eqnarray} 
Figure \ref{fig5} shows the transmission probability $|T_{-1}|^2$ of the quadrupolar mode, when the barrier potential is given by Eq.~(\ref{eq.8}). In contrast to the Bogoliubov mode and the transverse spin wave (which are described by Eqs. (\ref{eq.12}) and (\ref{eq.13}), respectively), the interaction effects on the quadrupolar mode cannot be simply described by the single parameter $c_+=c_1+c_2$. Because of this, the coupling $c_1$ remains in the dimensionless equation (\ref{eq.36}). (Note that $c_1$ in Eq.~(\ref{eq.36}) actually denotes ${\bar c}_1=c_1/c_+$.) Indeed, we find in Fig.~\ref{fig5} that the transmission probability strongly depends on the value of $c_1$. Although the quadrupolar spin mode could show perfect transmission at a certain value of $k$ (See, for example, Fig.~\ref{fig5} (a)), it owes to a fine tuning of the parameter. The tunneling property of this mode is rather close to the ordinary quantum mechanical tunneling of a particle in the sense that perfect reflection occurs when $k \to 0$~\cite{WatabeKato2010,WatabeDthesis,WatabeKatoLett,WatabeKatoFull}. 

\par
\section{Effects of spin-dependent barrier}\label{SecV} 
\par
So far, we have considered the case when a barrier potential couples to the local density of particles as $V(x)\rho(x)$. Since the transverse spin wave is a Nambu-Goldstone mode associated with spontaneous breakdown of rotational spin symmetry, the tunneling property of this mode is expected to be strongly affected by a spin-dependent magnetic barrier. To see this, we examine tunneling through a magnetic potential barrier
\begin{equation} 
V_{\rm s} (x)=-J(x)F_z(x).
\label{eq.38}
\end{equation}
Although this tunneling problem is somehow academic in cold atom gases, it is still instructive to see how the perfect transmission of the transverse spin wave is related to the symmetries broken or preserved by the barrier. For simplicity, we assume the absence of supercurrent ($q=0$), as well as the external magnetic field ($B=0$). We also take $J(x)>0$, which enhances the $S_z=+1$ spin component near the barrier. 
When we replace $V(x)\rho(x)$ in Eq.~(\ref{eq.2b}) by $V_{\rm s}(x)$, we again obtain a $3\times 3$-matrix GP equation similar to Eq.~(\ref{eq.5}). By numerically solving the GP equation with the boundary condition $(\Phi_{+1},\Phi_{0},\Phi_{-1}) = (\sqrt{\rho_0},0,0)$ for $|x| \gg 1$, we have confirmed that $\Phi_{0} = \Phi_{-1} = 0$ even near the magnetic potential 
$V_{\rm s}(x)$. Thus, one may consider the single-component GP equation for $\Phi_{+1}(x)$, given by
\begin{align}
0 = 
\left [
-\frac{1}{2} \frac{d^{2}}{dx^{2}}
- J (x)
- 1 
+ \rho(x)
\right ] \Phi_{+1} (x). 
\label{eq.39}
\end{align} 
\par
Equations for the collective modes can be also obtained in the same manner as we have done in Sec.~II. The results are
\begin{align} 
i {\partial \phi_{+1} \over \partial t}
 = &
\left [
-{1 \over 2} \frac{d^{2}}{dx^{2}}
- J (x)- 1 +2 \rho(x) 
\right ] \phi_{+1}+\Phi_{+1}^2\phi_{+1}^*,
\label{eq.40}
\\
i {\partial {\phi}_{0} \over \partial t}
= &  
\left [
-\frac{1}{2} \frac{d^{2}}{dx^{2}} 
- 1 
+ \rho(x)
\right ]  {\phi}_{0}, 
\label{eq.41}
\\
i {\partial {\phi}_{-1} \over \partial t}
= &
\left [
-\frac{1}{2} \frac{d^{2}}{dx^{2}}
+ J (x)
-
1
+ 
(1 - 2c_{1})
\rho(x) 
\right ]  {\phi}_{-1}. 
\label{eq.42}
\end{align}
\par
Equations (\ref{eq.39}) and (\ref{eq.40}) have the same 
forms as the GP equation and the Bogoliubov equation in the scalar BEC, respectively. Thus, the Bogoliubov mode exhibits perfect transmission through the magnetic barrier $V_{\rm s} (x)$ in the low energy limit.  
\par
To see whether or not the transverse spin wave shows perfect transmission, it is convenient to consider the $\delta$-functional barrier $J(x)=J_0\delta(x)$ ($J_0>0$). In this simple case, the GP equation (\ref{eq.39}) has the exact solution 
\begin{equation}
\Phi_{+1} (x) =  \coth{(|x|+X_{0})}, 
\label{eq.43}
\end{equation}
where $X_{0}=\coth^{-1}[(\sqrt{J_0^2 + 4}-J_0)/2]$. Substituting $\rho(x)=|\Phi_{+1}(x)|^2$ into Eq.~(\ref{eq.41}), one finds that the transverse spin wave solution has the form
\begin{eqnarray}
\begin{array}{ll}
{\phi}_{0} (x) =  f_{-}(k,x) + R_0 f_{-}(- k,x), & ~~~(x \leq 0),\\
{\phi}_{0} (x) = T_0 f_{+}(k,x), &~~~(x \geq 0),\\ 
\end{array}
\label{eq.44}
\end{eqnarray}
where $f_{\pm} (k,x) = [\mp ik + \coth (\pm x +X_{0})] e^{ikx}$. Since the potential barrier $J(x)$ is absent in Eq.~(\ref{eq.41}), the boundary conditions at $x=0$ are simply given by the matching of the wave function and its gradient, 
which give
\begin{align}
T_0 = 
& 
- \frac{k (1 + k^{2})}{(k + i \Gamma ) (- 1 + k^{2} + ik \Gamma + \Gamma^{2})}, 
\label{eq.45}
\\
R_0 = & 
\frac{- i\Gamma (1 - \Gamma^{2})}{(k + i \Gamma) (- 1 + k^{2} + ik \Gamma + \Gamma^{2})}, 
\label{eq.46}
\end{align}
where $\Gamma=\coth(X_0)$. The coefficient $T_0$ in 
Eq.~(\ref{eq.46}) vanishes in the limit $k\to 0$, unless $\Gamma=1$ (i.e., $J_0=0$). Thus, in contrast to the `non-magnetic' potential $V(x)$ discussed in Sec.~III, the magnetic barrier $V_{\rm s}(x)$ leads to perfect {\it reflection} of the transverse spin wave in the low energy limit. 

\begin{figure}[tbp]
\begin{center}
\includegraphics[width=6cm]{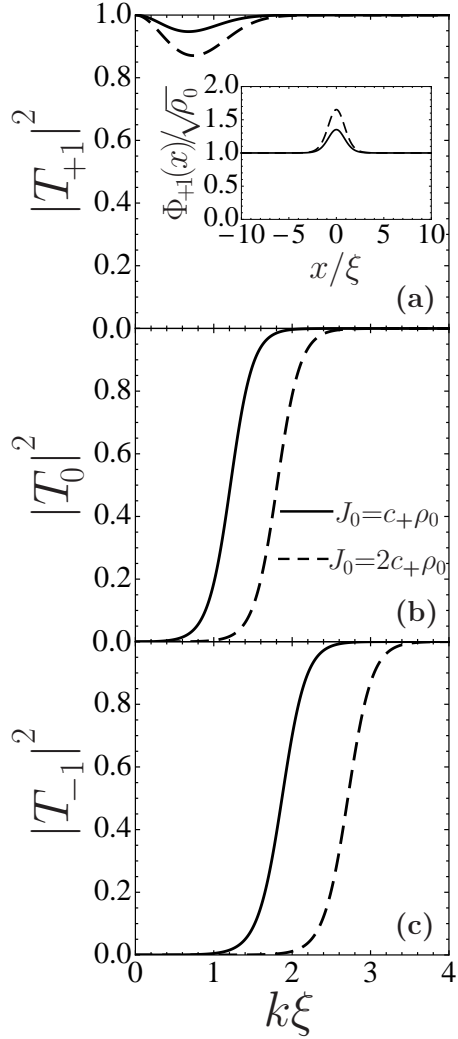}
\end{center}
\caption{
Transmission probability through the magnetic potential barrier when $J(x)=J_0e^{-x^2/\xi^2}$. (a) Bogoliubov mode. (b) transverse spin wave. (c) quadrupolar spin wave. The inset shows the spatial variation of the ferromagnetic condensate wave function $\Phi_{+1}(x)$.
}
\label{fig6}
\end{figure} 

\par
Figure \ref{fig6} shows the transmission probabilities of the three collective excitations through the magnetic barrier potential. As discussed above, the Bogoliubov mode shown in panel (a) only exhibits the perfect transmission at $k=0$. Although the transverse spin wave does not feel the magnetic potential (See Eq.~(\ref{eq.41})), it is still scattered by the `barrier' formed by the condensate wave function around $x=0$ shown in the inset of panel (a). Since Eq.~(\ref{eq.41}) has the same form as the ordinary one-particle Schr\"odinger equation with the `potential barrier' $|\Phi_{+1}(x)|^2-1$, the perfect reflection at $k=0$ can be understood as a usual  property in one-particle quantum mechanics.
\par
%
\begin{figure}[tbp]
\begin{center}
\includegraphics[width=6cm]{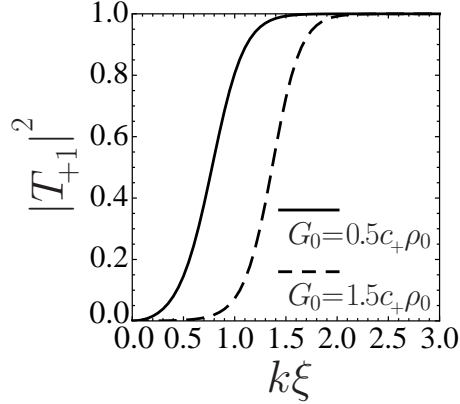}
\end{center}
\caption{
Transmission probability of the Bogoliubov mode through the barrier given by Eq.~(\ref{eq.47}). We take $G(x)=G_0e^{-x^2/\xi^2}$.
}
\label{fig7}
\end{figure} 
\par

As we have seen, the spin-dependent barrier suppresses the perfect transmission of the transverse spin wave but the Bogoliubov mode still exhibits anomalous tunneling. This fact implies that the perfect transmission (or the anomalous tunneling phenomenon) of a Nambu-Goldstone mode is strongly affected by the barrier that breaks the symmetry related to the origin of the mode. 
To see this in a simple manner, we replace $V(x) \rho(x)$ in Eq.~(\ref{eq.2b}) by the one that couples to the 
local phase 
$\varphi_{+1}(x)$ of the condensate wave function $\Phi_{+1}$~\cite{note2}, such that 
\begin{equation}
V_{\rm G} (x)=G(x)\frac{\Phi_{+1} (x) + \Phi_{+1}^{*} (x)}{|\Phi_{+1} (x) |}
    =2 G(x)\cos\varphi_{+1}(x).
\label{eq.47}
\end{equation}
When $G(x)=G_0\delta(x)$, the transmission probability $|T_{+1}|^2$ of the Bogoliubov mode with the incident momentum $k$ is given by
\begin{equation}
|T_{+1}|^2={\eta^2 \over \eta^2+G_0^2\kappa^6}.
\label{eq.48}
\end{equation}
(We summarize the derivation in Appendix C.) Here, $\eta=k[|G_0|k^2+\kappa(k^2+\kappa^2)]$, and $\kappa=\sqrt{2(\sqrt{1+E^2}+1)}$, where $E=(k/\sqrt{2})\sqrt{(k/2)^2+2}$ is the mode energy.  Equation (\ref{eq.48}) vanishes in the low energy limit $E\rightarrow 0$, so that perfect {\it reflection} occurs. Even for a barrier with a finite width, this conclusion is unaltered, as shown in Fig.~\ref{fig7}. (Equations to obtain this figure are summarized in Appendix C.) The above two results for the transverse spin wave and the Bogoliubov mode indicate that the anomalous tunneling phenomenon is deeply related to 
the invariance of the barrier potential with respect to continuous symmetries that are broken spontaneously, such as the U(1) gauge and spin rotation. 

\section{Summaries and Discussions}\label{SecVI}

To summarize, we have investigated tunneling properties of the collective Bogoliubov mode, as well as spin wave excitations (transverse spin wave and quadrupolar mode), in the ferromagnetic state of a spin-1 spinor Bose-Einstein condensate. Within the framework of the mean-field theory at $T=0$, we have calculated the tunneling probability across a barrier potential of each mode as a function of the incident momentum, to see whether or not anomalous tunneling phenomenon occurs.
\par
We showed that, as in the case of a scalar BEC, the Bogoliubov mode 
also exhibits the anomalous tunneling phenomenon 
in the ferromagnetic state of a spin-1 spinor BEC in the presence of a barrier that couples only to local density. 
Namely, perfect transmission always occurs in the low energy limit, even in the presence of supercurrent, unless the superflow is in the critical supercurrent state. 
In the critical supercurrent state, tunneling of the Bogoliubov mode is accompanied by a finite reflection. All these results are the same as the case of a scalar BEC.
\par
The transverse spin wave, which is the other Nambu-Goldstone mode associated with the broken spin rotational symmetry, 
also shows the anomalous tunneling phenomenon in the presence of a barrier that couples only to local density. 
However, while this phenomenon occurs in the long wavelength limit in the case of the Bogoliubov mode, perfect transmission of the transverse spin wave is realized when 
the momentum of the spin wave coincides with that of supercurrent. 
In addition, this phenomenon still occurs in the critical supercurrent state, which is also to be contrasted with the case of the Bogoliubov mode. 
On the other hand, when perfect transmission occurs, the wave function of the transverse spin wave has the same form as the condensate wave function in the current carrying state. 
This property is shared with the Bogoliubov mode in the long wavelength limit. 
\par
This result is important in the context described below.
The perfect transmission of the transverse spin wave in the current-flowing state provides much simpler example of anomalous tunneling, compared to earlier cases. 
In the earlier cases (the Bogoliubov mode~\cite{Kovrizhin2001,Kagan2003,Danshita2006,Kato2008,Watabe2008,Tsuchiya2008,Ohashi2008,Watabe2009RefleRefra,Takahashi2009,Tsuchiya2009,Takahashi2010}, 
the transverse spin wave in the absence of supercurrent~\cite{WatabeKato2010,WatabeDthesis,WatabeKatoLett,WatabeKatoFull}), 
the anomalous tunneling always occurs in the limit where the wavenumber $k$ approaches zero. In the present case, on the other hand, anomalous tunneling occurs at a finite wavenumber. 
When we discuss the tunneling problem in the limit $k\rightarrow 0$, we have to calculate the transmission rate for finite $k$ first, and then consider the limit $k\rightarrow 0$. 
This limiting procedure is unnecessary when we consider the tunneling problem of particles with finite wavenumber. 
This is the reason why the proof of perfect transmission for $q\ne 0$ presented in Sec.~II is simpler than those for $q=0$ given in \cite{WatabeKatoLett} and Appendix A in the present paper.
Owing to this simplicity of the perfect transmission of the spin wave in the current-flowing state, we can prove the perfect transmission of the spin wave for a general potential barrier and $q\ne 0$ as a direct consequence of the coincidence of wave functions between the spin wave and the flowing condensate. 
\par
We have also examined tunneling properties of the quadrupolar mode. This collective mode has a finite excitation gap, and is not the Nambu-Goldstone mode. In contrast to the above two Nambu-Goldstone modes, the quadrupolar mode does not exhibit the anomalous tunneling behavior. 
\par 
The tunneling properties of excitations are determined by the properties of potentials as well as characters of excitations. 
When a barrier potential locally destroys a symmetry that is spontaneously broken in the spinor BEC, the anomalous tunneling phenomenon of the corresponding Nambu-Goldstone mode disappears. 
Indeed, perfect transmission of the transverse spin wave does not occur, when the barrier potential depends on spin and locally breaks spin rotation symmetry. 
Similarly, perfect transmission of the Bogoliubov mode does not occur either, when the barrier potential depends on the local phase of the condensate wave function and locally breaks U(1) gauge symmetry. 
The presence of the symmetry breaking potential inhibits the wave function of a Nambu-Goldstone mode from coinciding with the condensate wave function. 
These results indicate that the symmetry that the barrier potential possesses is crucial for the anomalous tunneling phenomena of Nambu-Goldstone modes. 
We infer that anomalous tunneling occurs in a Nambu-Goldstone mode only in the presence of symmetry-preserving barrier potentials. 
This result shown in Sec.~\ref{SecV} is significant because this is the first study on relation between anomalous tunneling and symmetry-property of potentials. 
\par 
The present paper has treated the tunneling problem for all the three excitations separately. 
It seems to be difficult to create an excitation as a single mode from the spin-1 BEC to examine tunneling properties of excitations experimentally; 
but, this difficulty can be overcome by making use of the Bragg scattering and a strong magnetic field~\cite{WatabeKatoFull}. 
The Bragg scattering can selectively create an excitation with a certain momentum $p$ and a certain energy $E$. 
When energies of the three collective excitations at small momenta are well separated from each other by using the Zeeman effect under a strong magnetic field, 
one would be able to selectively create one of the three excitations as a single mode. 
\par 
Since low energy properties of the ferromagnetic spinor BEC are dominated by the three collective excitations, 
our results would be useful for understanding dynamical properties in this superfluid phase. 
\par
In this paper, we have focused on the ferromagnetic phase of a spin-1 BEC. 
It is an interesting problem to extend our present work to the polar phase, which is the other phase of the spin-1 BEC. 
We have examined the same problem for the spin wave modes in the current carrying state of the polar phase~\cite{WKO}. 
We found that the anomalous tunneling phenomenon of the spin wave modes occurs unless the system is in the critical current state. 
The anomalous tunneling phenomenon in itself is the same character as the Bogoliubov mode and the transverse spin mode in the ferromagnetic phase. 
However, in the critical current state, the tunneling property is qualitatively changed by the interaction parameter. 
These findings will be reported in a separate paper~\cite{WKO}. 
\acknowledgements

S.W. and Y.O. would like to thank D. Inotani, T. Kashimura, and R. Watanabe for discussions. This work was supported by Grant-in-Aid for Scientific Research 
(20500044, 21540352, 22540412) from JSPS, Japan.



\appendix
\section{Proof of perfect transmission of transverse spin wave for a generic barrier at $q=0$}
The proof of perfect transmission of the transverse spin wave for $q\ne0$ is presented in Sec.~\ref{SecIII} but this proof is not applicable when $q=0$. While the proof for the case with $q=0$ has been given in \cite{WatabeKatoLett}, we present in this appendix another proof of perfect transmission of the transverse spin wave at $q=0$ in order to show that the similarity between the wave function of an excitation and the supercurrent wave function leads to perfect transmission for the case with $q=0$ as well as the case with $q\ne 0$. In this appendix, we abbreviate $\Phi_{+1}(x,q)$ as $\Phi(x,q)$ for simplicity.

First we introduce
\begin{equation}
f_q(x)=|\Phi(x,q)|^2-|\Phi(x,0)|^2,
\end{equation}
which is ${\cal O}(q^2)$ because $|\Phi(x,q)|^2$ depends on $q$ only through $q^2$ (Equation for $|\Phi(x,q)|^2$ contains $q$ only through $q^2$ ). When $q=0$, the equation of motion for the transverse spin wave with the momentum $k>0$
\begin{equation}
\left[-\frac12\frac{d^2}{dx^2}+V(x) - 1 - \frac{k^2}{2}+|\Phi(x,0)|^2\right]\phi_0(x)=0
\label{eq: eom-tsw-q=0}
\end{equation}
is rewritten as 
\begin{equation}
\left[\hat{h}_k-f_k(x)\right]\phi_0(x)=0
\label{eq: q0k}
\end{equation}
with 
\begin{equation}
\hat{h}_k\equiv -\frac12\frac{d^2}{dx^2}+V(x) -1-\frac{k^2}{2}+|\Phi(x,k)|^2. 
\end{equation}
We seek for the solution to (\ref{eq: eom-tsw-q=0}) satisfying the boundary condition  Eq.~(\ref{eq.32}) for the tunneling problem, assuming the form
\begin{equation}
\phi_0(x)=N_k \left[\Phi(x,k)+\Delta\phi_k(x)\right]
\label{eq: phi0}
\end{equation}
with a normalization factor $N_k$. Equation (\ref{eq: q0k}) is rewritten as
\begin{equation}
\hat{h}_k \Delta\phi_k(x)=N_k^{-1} f_k(x)\phi_0(x).
\label{eq: q0k-1}
\end{equation}
Substituting the following expressions:
\begin{eqnarray}
&&\Delta\phi_k(x)=\chi_k(x)\Phi(x,k)\label{eq: chiPsi}\\
&&\chi_k(x)=2N_k^{-1}\int_0^{x}dx'[\Phi(x',k)]^{-2}\int_{x'}^{\infty}dx'' f_k(x'')\phi_0(x'')\Phi(x'',k)\label{eq: chi}
\end{eqnarray}
into (\ref{eq: q0k-1}), we see that (\ref{eq: chiPsi}) with  (\ref{eq: chi}) yields a solution that we seek for. However, it is a formal solution; Equation~(\ref{eq: chi}) contains $\phi_0$ in the right-hand side.
In the following,  we consider the asymptotic form of $\chi_k(x)$ at $|x|\gg 1$.
We first note that $|\Phi(x\rightarrow \pm \infty,k)|^2\rightarrow 1$ irrespective of $k$ and 
hence $f_k$ is localized around $x=0$.
We then see that
\begin{equation}
\int_{x'}^{\infty}dx'' f_k(x'')\phi_0(x'')\Phi(x'',k)
\label{fphiPhi}
\end{equation}
approaches zero when $x'\rightarrow \infty$ while it approaches a constant
\begin{equation}
\lambda_k\equiv \int_{-\infty}^{\infty}dx'' f_k(x'')\phi_0(x'')\Phi(x'',k)
\end{equation}
when $x'\rightarrow -\infty$. 
From this property and the asymptotic behavior 
$\Phi(x\rightarrow \pm \infty,k)\rightarrow e^{i [k x + \Delta\varphi_k{\rm sgn}(x)/2]}$ of the condensate wave function, it follows that 
\begin{equation}
N_k \chi_k(x)\rightarrow
\left\{
\begin{array}{cc}
a_k,& (x\gg 1), \\
-\lambda_k(e^{-2i k x }-1)e^{i \Delta\varphi_k}/(i k)+b_k,& (x\ll -1).
\end{array}
\right.\label{eq: chiasymptotic}
\end{equation}
Here and in the following, we make $k$-dependence of the phase shift $\Delta\varphi$ explicit. 
The constants $a_k$, $b_k$ are, respectively, given by
\begin{eqnarray}
a_k&\equiv&2\int_0^\infty dx'[\Phi_k(x')]^{-2}\int_{x'}^\infty dx'' f_k(x'')\phi_0(x'')\Phi(x'',k),\nonumber\\
b_k&\equiv&-2\int_{-\infty}^0dx' \left[\Phi(x',k)^{-2}-e^{i ( -2 k x' + \Delta\varphi_k )}\right]\int_{x'}^\infty dx'' f_k(x'')\phi_0(x'')\Phi(x'',k)\nonumber\\
& &+2\int_{-\infty}^0 dx' e^{i( - 2 k x' + \Delta\varphi_k) }\int_{-\infty}^{x'}dx'' f_k(x'') \phi_0(x'') \Phi(x'', k).
\end{eqnarray}
Substituting (\ref{eq: chiPsi}) with (\ref{eq: chi}) into (\ref{eq: phi0}) and 
using (\ref{eq: chiasymptotic}), we obtain
\begin{equation}
\phi_0(x)\rightarrow
\left\{
\begin{array}{cc}
(N_k+a_k)e^{i \Delta\varphi_k/2}e^{i k x }, & (x\gg 1),\\
\left [ N_k-i (\lambda_k/k)e^{i \Delta\varphi_k}+b_k \right] e^{-i \Delta\varphi_k/2}e^{i k x}&\\
+i(\lambda_k/k)e^{i\Delta\varphi_k/2}e^{-i k x }, &  (x\ll -1).
\end{array}
\right.\label{eq: phi0asymptotic}
\end{equation}
From (\ref{eq: phi0asymptotic}) and (\ref{eq.32}), the solution with 
\begin{eqnarray}
&&N_k=e^{i \Delta\varphi_k/2}-b_k+i (\lambda_k/k)e^{i \Delta\varphi_k}, \\
&&T_0=e^{i \Delta\varphi_k}+i (\lambda_k/k)e^{i 3\Delta\varphi_k/2}+(a_k-b_k)e^{i \Delta\varphi_k/2}, \\
&&R_0=i  (\lambda_k/k)e^{i \Delta\varphi_k/2}
\end{eqnarray}
follows. 

As noted previously, $f_k={\cal O}(k^2)$ when $|k|\ll 1$. $\phi_0(x)$ is at most ${\cal O}(1)$ (when the perfect reflection occurs in the limit $k\rightarrow 0$, the wave function is ${\cal O}(k)$) and $\Phi(x,k)$ is ${\cal O}(1)$. The quantities $a_k$, $b_k$, $\lambda_k$ contain commonly $f_k\phi_0 \Phi(x,k)$ as a factor in their integrands and hence they are ${\cal O}(k^2)$. The phase shift $\Delta\varphi_k$ is ${\cal O}(k)$ for $|k|\ll 1$. We thus obtain
\begin{equation}
T_0=1+i k\lim_{k\rightarrow 0}\left(\Delta\varphi_k/k +\lambda_k/k^2\right)+{\cal O}(k^2),
\end{equation}
\begin{equation}
R_0=i k \lim_{k\rightarrow 0}(\lambda_k/k^2)+{\cal O}(k^2)
\end{equation}
and the perfect transmission $|T_0|^2\rightarrow 1$ occurs in the limit $k\rightarrow 0$. 

\section{transverse spin wave solution in the low energy limit in the absence of supercurrent}
In this appendix, we show that the transverse spin wave mode has the same form as the current carrying condensate wave function in the long wavelength limit, even when the supercurrent is absent ($q=0$) and the potential barrier is given by $V(x)=V_0\delta(x)$.
\par
Assuming that the supercurrent momentum $q$ is small, we expand the condensate wave function in Eq.~(\ref{eq.17}) in terms of $q$ 
up to 
${\mathcal O} (q)$, which gives
\begin{align}
\Phi_{+1} (x) 
=  e^{i [ qx - {\rm sgn}(x) \theta_{q}]} 
[{\bar \gamma} (x) - iq {\rm sgn} (x)]\equiv \Phi_{+1}(x,q).
\label{eq.a1}
\end{align} 
Here, ${\bar \gamma}(x)=\tanh (|x| + x_{0})$, and $\theta_{q}=-q/{\bar \gamma}(0)$. We also expand the spin wave solution in Eq.~(\ref{eq.21}) around $k=0$ when $q=0$. The amplitude transmission and reflection coefficients in Eqs. (\ref{eq.23}) and (\ref{eq.24}) are expanded as, respectively,
\begin{align}
T_{0} = & 1 + ik \left ( {\bar \gamma}(0) + \frac{1}{{\bar \gamma}(0)} \right ), 
\label{eq.a2}
\\
R_{0} = & i k \left ( {\bar \gamma}(0) - \frac{1}{{\bar \gamma}(0)} \right ).
\label{eq.a3b}
\end{align} 
We then have, within the accuracy of ${\mathcal O} (k)$, 
\begin{eqnarray}
{\phi}_{0} (x\le 0) 
&=&  
[ + ik + {\bar \gamma} (x) ] e^{ikx} + i k 
\left ( {\bar \gamma}(0) - \frac{1}{{\bar \gamma(0)}} \right ) 
[- ik  + {\bar \gamma} (x) ] e^{-ikx} 
\nonumber
\\
&\simeq& 
e^{ikx} 
\left \{ 
\left [ 
1 + i k \left ( {\bar \gamma}(0) - \frac{1}{{\bar \gamma}(0)} \right )
\right ] 
{\bar \gamma} (x) 
+ i k 
\right \} 
\nonumber
\\
& \simeq & 
e^{ik{\bar \gamma}(0)}
e^{ikx} 
e^{- i k / {\bar \gamma}(0) } \left [ {\bar \gamma} (x) + i k  \right ], 
\label{eq.a3}
\end{eqnarray} 
and
\begin{align} 
{\phi}_{0} (x\ge 0) = 
& 
[- i k + {\bar \gamma} (x) ] e^{ikx} \left [ 1 + ik \left ( {\bar \gamma}(0) + \frac{1}{{\bar \gamma}(0)} \right ) \right ] 
\nonumber
\\
\simeq  & 
e^{ik{\bar \gamma}(0)}
e^{ikx} 
e^{ + i k / {\bar \gamma}(0) } \left [ {\bar \gamma} (x) - i k \right ]. 
\label{eq.a4}
\end{align}
Comparing Eqs. (\ref{eq.a3}) and (\ref{eq.a4}) with Eq.~(\ref{eq.a1}), we have
\begin{align}
{\phi}_{0} (x) 
= & 
e^{ik{\bar \gamma}(0)}
e^{ i [ kx - {\rm sgn} (x) \theta_{k} ]} \left [ {\bar \gamma} (x) - i k {\rm sgn} (x) \right ]+{\cal O}(k^{2}) 
\nonumber
\\
= & 
e^{ik{\bar \gamma}(0)} \Phi_{+1} (x,k)+{\cal O}(k^{2}). 
\label{eq.a5}
\end{align} 
Thus, the wave function of the transverse spin wave in the long wavelength limit ($k\to 0$) has the same form as the condensate wave function in the supercurrent state with the momentum $k$ within the accuracy of ${\mathcal O} (k)$.

\par
\section{Tunneling through a barrier coupling to the phase of condensate}
\par
In the case of Eq.~(\ref{eq.47}), the GP equation is given by
\begin{align}
\left [ 
- \frac{1}{2} \frac{d^{2}}{dx^{2}} - 1 + \rho(x)+ 
G (x)
\frac{\Phi^{*}_{+1} (x) - \Phi_{+1} (x)}{2 |\Phi_{+1} (x)|^{3}} 
\right ] \Phi_{+1} (x) = 0. 
\label{eq.a6}
\end{align} 
In this case, when $G (x) < 0$ ($G (x) >0$), the uniform solution $\Phi_{+1}(x)=1$ ($\Phi_{+1}=-1$) gives the minimum energy. 
When $G(x) = G_{0} \delta (x)$, the Bogoliubov equation for the Bogoliubov mode is then given by
\begin{equation}
i {\partial \phi_{+1}(x,t) \over \partial t}=
\left (
-{\displaystyle \frac{1}{2} \frac{d^{2}}{dx^{2}}} + 1
\right ) 
\phi_{+1}(x,t)
-\phi_{+1}^*(x,t)
+{\displaystyle \frac{1}{2}} |G_0 | \delta (x)
[\phi_{+1}(x,t)+\phi_{+1}^*(x,t)].
\label{eq.a7}
\end{equation}
As usual, the last term in Eq.~(\ref{eq.a7}) may be conveniently absorbed into the boundary conditions at $x=0$. Setting $\phi_{+1}(x,t)=u(x) e^{-iE t} - v^*(x) e^{i E t}$, one obtains an eigenvalue problem
\begin{align}
E 
\begin{pmatrix}
u(x) \\ v(x)
\end{pmatrix} 
= 
\begin{pmatrix}
\displaystyle{ -\frac12\frac{d^2}{dx^2}}+1  & - 1 \\
1 & \displaystyle{\frac12\frac{d^2}{dx^2}}-1
\end{pmatrix} 
\begin{pmatrix}
u(x) \\ v(x)
\end{pmatrix}.
\label{eq.a8}
\end{align}
The basis functions of general solutions to (\ref{eq.a8}) are given by 
\begin{equation}
\begin{pmatrix}
u(x) \\ v(x)
\end{pmatrix}
=  
\begin{pmatrix}
A \\ B 
\end{pmatrix} 
e^{\pm i k x},\quad
\begin{pmatrix}
B \\ - A
\end{pmatrix} 
e^{\pm \kappa x}. 
\end{equation}
Here we introduce $k=\sqrt{2 ( \sqrt{ 1 + E^{2} } - 1 )}$, $\kappa= \sqrt{2 ( \sqrt{ 1 + E^{2} } + 1 )}$, and   
\begin{align}
A= \sqrt{ \frac{1}{2} \left (  \frac{\sqrt{1 + E^{2} }}{E} + 1 \right )}, 
\qquad 
B= \sqrt{ \frac{1}{2} \left (  \frac{\sqrt{1 + E^{2} }}{E} - 1 \right )}, 
\label{eq.a10}
\end{align}
which are normalized so as to satisfy $A^{2} - B^{2} = 1$.

Under the condition that the Bogoliubov mode is injected from $x=-\infty$, we set 
\begin{align}
\begin{pmatrix}
u(x) \\ v(x)
\end{pmatrix}
= & 
\begin{pmatrix}
A \\ B 
\end{pmatrix} 
e^{+ i k x} 
+ 
R_{+1} 
\begin{pmatrix}
A \\ B 
\end{pmatrix} 
e^{- i k x} 
+ a
\begin{pmatrix}
B \\ - A
\end{pmatrix} 
e^{\kappa x}, 
&
(x \leq 0),  
\label{eq.a9}
\\
\begin{pmatrix}
u(x) \\ v(x)
\end{pmatrix}
= & 
T_{+1}
\begin{pmatrix}
A \\ B
\end{pmatrix} 
e^{ i k x} 
+ b
\begin{pmatrix}
B  \\ - A 
\end{pmatrix} 
e^{- \kappa x}, 
&
(x \geq 0). 
\label{eq.a9b}
\end{align}
The boundary conditions at $x=0$ are the matching of Eqs. (\ref{eq.a9}) and (\ref{eq.a9b}), and 
\begin{eqnarray}
{d \over dx}
\left(
\begin{array}{c}
u(x) \\ v(x)
\end{array}
\right)_{x=+0}
-
{d \over dx}
\left(
\begin{array}{c}
u(x) \\ v(x) 
\end{array}
\right)_{x=-0}
=|G_0|
\left(
\begin{array}{c}
u(0)+v(0) \\ u(0)+v(0) 
\end{array}
\right). 
\label{eq.a10b}
\end{eqnarray} 
The coefficients in Eqs. (\ref{eq.a9}) and (\ref{eq.a9b}) are determined from these boundary conditions as
\begin{align}
T_{+1} = & \frac{\eta}{\eta + i |G_{0}| \kappa^{3}}, 
\label{eq.a11}
\\ 
R_{+1} = & \frac{-i |G_{0}| \kappa^{3}}{\eta + i |G_0| \kappa^{3}}, 
\label{eq.a12}
\\
a =& b 
 = \frac{2 E k |V_{0}^{\rm (c)} | }{\eta + i |G_0 | \kappa^{3}}, 
\label{eq.a13}
\end{align} 
where $\eta \equiv k[ |G_0| k^{2} + \kappa (k^{2} + \kappa^{2}) ]$. The transmission probability (Eq.~(\ref{eq.48})) follows from Eq.~(\ref{eq.a11}).
\par
When the barrier has a finite width (e.g., $G(x)=G_0e^{-x^2/\xi^2}$), the Bogoliubov equation, corresponding to Eq.~(\ref{eq.a8}), becomes 
\begin{align}
E 
\begin{pmatrix}
u(x) \\ v(x)
\end{pmatrix} 
= 
\begin{pmatrix}
- 
{\displaystyle \frac{1}{2} \frac{d^{2}}{dx^{2}} }
- 1 + 2\rho(x) + \chi(x) &  - \Phi^{2}(x) - W(x) \\
\Phi^{* 2}(x) + W^{*}(x)  & - \left [- {\displaystyle \frac{1}{2} \frac{d^{2}}{dx^{2}} } - 1 + 2\rho(x) \right ] - \chi^{*}(x)
\end{pmatrix} 
\begin{pmatrix}
u(x) \\ v(x)
\end{pmatrix}, 
\label{eq.a14}
\end{align}
where 
\begin{align}
\chi (x) \equiv & \frac{1}{2} G(x) \frac{1}{|\Phi (x)|^{3}} 
\left  [ \Phi^{*} (x) - 2 \Phi (x)  - \frac{3}{2} \frac{|\Phi (x) |^{2} - \Phi^{2} (x)}{\Phi (x) } \right ] , 
\label{eq.a15}
\\
W (x) \equiv & \frac{1}{2} G(x) \frac{1}{|\Phi (x)|^{3}} 
\left [ \Phi (x) - \frac{3}{2} \frac{|\Phi (x) |^{2} - \Phi^{2} (x)}{\Phi^{*} (x) } \right ] .  
\label{eq.a16}
\end{align}
In this case, we numerically solve Eq.~(\ref{eq.a14}), regarding Eqs. (\ref{eq.a9}) and (\ref{eq.a9b}) as the boundary conditions at $x=-\infty$ and $x=\infty$, respectively. 

Before closing this appendix, we briefly remark on transmission of the transverse spin mode through the potential (\ref{eq.47}), 
which couples to the local phase of the condensate wave function $\Phi_{+1}$, and gives the uniform solution of $\Phi_{+1}$. 
In this case, the equation of the excitation for $S_{z} = 0$ is given by (\ref{eq.13}), where $\rho (x)$ is spatially constant (i.e., $\rho (x) = \rho_{0}$) and $V(x) =0$. 
(The potential (\ref{eq.47}) does not enter in (\ref{eq.13}), since it does not couple to the component $S_{z} = 0$). 
As a result, the transverse spin mode propagates through a uniform medium, 
and shows perfect transmission. 



\begin{thebibliography}{99}
\bibitem{note} The Bogoliubov equations usually mean the coupled equations which are obtained from Eq.~(\ref{eq.12}) by executing the Bogoliubov transformation. In this paper, however, we simply call Eqs. (\ref{eq.12})-(\ref{eq.14}) the Bogoliubov equations describing collective excitations in the spinor BEC.
\bibitem{NoteFunc} We suppose that the barrier has even symmetry, so that the asymptotic behavior of the wave function is written in this simple form.
\bibitem{note2} We do not discuss the reality of this potential here. 
We only use it to confirm that the anomalous tunneling phenomenon of the Bogoliubov (Nambu-Goldstone) mode associated with the broken U(1) gauge symmetry is strongly affected when the barrier destroys the U(1) symmetry.
\end{thebibliography}
\end{document}